\documentclass[11pt,a4paper]{iopart}
\usepackage{graphicx}
\usepackage{cite}
\usepackage[export]{adjustbox}
\usepackage[scr=rsfs]{mathalpha}
\usepackage{textcomp,gensymb} 
\usepackage{amssymb}
\usepackage{amsfonts}
\usepackage{setspace}
\usepackage{amsmath}
\usepackage{iopams}
\usepackage{xcolor}

\def\bbq{\begin{equation}}
\def\ebq{\end{equation}}
\def\ba{\begin{array}}
\def\ea{\end{array}}
\def\n{\nonumber}

\def\la{\langle}
\def\ra{\rangle}

\begin{document}

\title[Stationary state of harmonic chains driven by boundary resetting]{Stationary state of harmonic chains driven by boundary resetting}

\author{Ritwick Sarkar and Pritam Roy}
\address{S. N. Bose National Centre for Basic Sciences, Kolkata 700106, India\\ E-mail: \texttt{ritwick.sarkar@bose.res.in}}

\begin{abstract}
We study the nonequilibrium steady state (NESS) of an ordered harmonic chain of $N$  oscillators connected to two walls which undergo diffusive motion with stochastic resetting. The intermittent resetting of the walls effectively emulates two nonequilibrium reservoirs that exert temporally correlated forces on the boundary oscillators. These reservoirs are characterized by the diffusion constants and resetting rates of the walls. We find that, for any finite $N$, the velocity distribution of the bulk oscillators remains non-Gaussian, as evidenced by a non-zero bulk kurtosis that decays $\sim N^{-1}$. We calculate the spatio-temporal correlation of the velocity of the oscillators $\la v_l(t) v_{l'}(t') \ra$ both analytically as well as using numerical simulation. The signature of the boundary resetting is present at the bulk in terms of the two-time velocity correlation of a single oscillator and the equal-time spatial velocity correlation. For the resetting driven chain, the two-time velocity correlation decay as $t^{-\frac{1}{2}}$ at the large time, and there exists a non-zero equal-time spatial velocity correlation $\la v_l(t) v_{l'}(t') \ra$ when $l \neq l'$. A non-zero average energy current will flow through the system when the boundary walls reset to their initial positions at different rates. This average energy current can be computed exactly in the thermodynamic limit. Numerically we show that the distribution of the instantaneous energy current at the boundary is independent of the system size. However, the distribution of the instantaneous energy current in the bulk approaches a stationary distribution in the thermodynamic limit.

\end{abstract}


\section{Introduction}\label{sec:intro}

It is worthwhile to understand the nonequilibrium steady state (NESS) of an extended system driven by equilibrium reservoirs to comprehend the transport properties of that system. A paradigmatic model of energy transport in one-dimensional systems was studied by Rieder, Lebowitz, and Lieb in 1967, which consists of a harmonic chain of $N$ oscillators, driven by two thermal reservoirs at the boundaries \cite{rllmothRieder}. For this model, the thermal driving at the two boundaries leads to a NESS with a nonzero average energy current flowing through the system. Several generalizations --- e.g.; inclusion of anharmonic interactions between the oscillators of the chain, pinning potential, and disorders,  were studied in the recent past. These studies conclude that these generalizations lead to some nontrivial behavior of the steady state including anomalous transport and nonlinear temperature profile \cite{advncpdhar,Transportbook,nakazawa,RoyDhar2008,Dhar2001,FPUT,FPUT_alternatingmass,
kundu_sanjib,kannan_12}.

The action of nonequilibrium reservoirs, which do not satisfy fluctuation-dissipation theorem (FDT)\cite{Kubo}, on single probe particles leads to unusual features like anomalous relaxation dynamics, negative viscosity, modification of equipartition theorem, etc \cite{ kafri2021,maggi2014,maes2020,active_bath,collapse_polymer, work_fluct,dissipation_activefluid, sup_diff_colloid,santra2022,bacterialbath2011,gopal2021}. Recently, the NESS of a harmonic chain driven by active baths (nonequilibrium bath) is studied in Ref.~\cite{activity_driven_chain,activity_stationary}. Here we model a nonequilibrium reservoir using stochastic resetting. Briefly, in our model, the boundary oscillators of a harmonic chain are attached to two walls that perform free diffusion with intermittent resetting. These walls act as nonequilibrium reservoirs at the two ends and the effect of these reservoirs on the NESS of the harmonic chain is studied in detail in this work.

Stochastic resetting refers to the process where the dynamics of a system stochastically stops and restarts from some predefined condition\cite{resetrev}. The paradigmatic example is where the position of a Brownian particle resets to a fixed point in space with some specific rate \cite{satyaprl_reset}. This leads to some striking features like a nonequilibrium stationary state, finite mean first-passage time, and anomalous relaxation behavior. The effect of stochastic resetting on well-known equilibrium and nonequilibrium processes has been observed theoretically \cite{ising,log_pot_reset,telegraphic,levyflight1,levyflight2,coagulation,particle_transport1, particle_transport2, particle_transport3, particle_transport4,rtp1,rtp2,gbp_reset,extreme_reset,work_fluctuation,fluct_interface1,fluct_interface2} as well as experimentally\cite{expt_reset1,expt_reset2,expt_reset3}. Stochastic resetting drives a system out-of-equilibrium irrespective of the underlying dynamics of the system. Stochastic resetting has found applications in the context of the search algorithm \cite{search1, search2, search3}, generation of controlled NESS \cite{controlled_NESS}, and quantum measurements \cite{measurement1,measurement2}. Moreover, stochastic resetting has been used to model a nonequilibrium bath \cite{noneq_bath} which consists of a collection of over-damped particles that are reset to a specific position.

In this paper, we introduce a simple model of a nonequilibrium reservoir using stochastic resetting dynamics. The action of tnonequilibrium reservoir is realized in terms of the wall that performs Brownian motion and stochastically resets to its initial position at a specific rate.  When the boundary oscillators of a one-dimensional harmonic chain are attached to two of these reservoirs, the chain attains NESS with an average nonzero energy current flowing through the system due to the action of these reservoirs. The average energy current can be easily calculated from the formalism used in Ref.~\cite{activity_driven_chain} and the value of the average energy current depends on the resetting rate of the boundary walls. We find that the velocity distribution of the bulk oscillator reaches the Gaussian distribution only at a thermodynamically large system size. The velocity distributions of the boundary oscillators are found to be size-independent with exponentially decaying tails. Numerical simulation results confirm that the kurtosis of the velocity of the bulk oscillators decays as $N^{-1}$. We also calculate the two-point spatio-temporal correlation of the velocity of the bulk oscillators. We find that at large time, the two-time velocity correlation $\la v_l(t) v_{l}(0) \ra$ decay as $t^{-\frac{1}{2}}$ and there exists a non-zero equal-time spatial velocity correlation $ \la v_l(t) v_{l'}(t) \ra$ in the bulk when $l\neq l'$. Instantaneous energy current distribution at the bulk reaches a stationary distribution in the thermodynamic limit. The exact expression of this stationary distribution is calculated in Ref.~\cite{activity_stationary}. In contrast, the distribution of instantaneous energy current at the boundary remains size-independent. We also discuss the effective thermal limit which is achieved at a high resetting rate.

The paper is organized as follows; in Sec.~\ref{sec_model}, we define the model and summarize our key findings. In Sec.~\ref{sec_vel_dist}, we discuss the velocity distribution and kurtosis profile of velocity.  The two-point spatio-temporal correlation of the velocity for the resetting driven chain is discussed in Sec.~\ref{sec_cross}. Sec.~\ref{sec_current} and Sec.~\ref{sec_thermal_case} are devoted to the distribution of instantaneous energy current and the effective thermal limit. We conclude in Sec.~\ref{sec_conclusion} along with some general remarks.

\section{Model and Results}\label{sec_model}
\begin{figure}[ht]
\centering

    \includegraphics[width=0.6\linewidth]{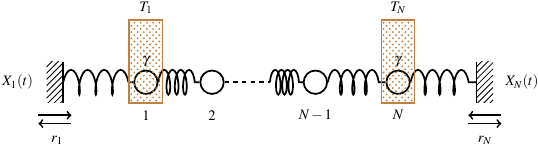} 

    \caption{ Schematic figure of a linear chain of oscillators connected to two walls that perform free diffusion and reset to the initial position with different rates. The boundary oscillators are also attached to two Langevin baths of temperature $T_1$ and $T_N$. }
    \label{fig:example}
\end{figure}
We consider a chain of $N$ oscillators, each of mass $m$, coupled to its nearest neighbors by springs of stiffness constant $k$.  Two boundary oscillators are in contact with Langevin heat baths at temperatures $T_1$ and $T_N$. These heat baths exert white noises as well as damping forces, proportional to the velocities, on the boundary oscillators. The boundary oscillators are also attached to two walls with springs of the same stiffness constant. These walls perform free diffusion and undergo intermittent resetting to their initial positions with different rates $r_1$ and $r_N$. We denote the displacement of the $l$-th oscillator from its equilibrium position as $x_l$ and the displacements of the left and right resetting walls as $X_1$ and $X_N$ respectively. These moving walls exert stochastic forces on the boundary oscillators due to the intermittent resetting dynamics. The equations of motion of the oscillators are,
\begin{equation}
m \ddot{x_l}=\begin{cases}
      -k(2x_1-x_2)+f_1-\gamma \dot{x}_1+\xi_1, & \text{when $l=1$, }\\
      -k(2x_l-x_{l-1}-x_{l+1}), & \text{$ \forall l\in [2,N-1] $, }\\
      -k(2x_N-x_{N-1})+f_N-\gamma \dot{x}_N+\xi_N. & \text{when $l=N$}.
      \end{cases}\label{eq:eom}
\end{equation}
$f_1=kX_1$ and  $f_N=kX_N$ are the stochastic forces exerted by the boundary walls. $\xi_1$ and $\xi_N$ are white noise from the Langevin heat bath acting on the boundary oscillators and satisfy the fluctuation-dissipation relation \cite{Kubo},
\begin{equation}
\la \xi_i(t) \xi_j(t') \ra=2 \gamma T_i \delta_{ij} \delta(t-t')\quad \text{and}~ i,j=1,N.
\end{equation}
$T_1$ and $T_N$ are the temperatures of the thermal reservoir at the two ends of the chain. $\gamma$ is the damping coefficient of the respective reservoir. Note that, we assume the motion of the wall is independent of the boundary oscillators.

The boundary walls perform free diffusion with Poisson resetting \cite{resetrev,satyaprl_reset}. In a small time interval $\Delta t$, the position of the $i$-th wall, relative to its equilibrium position, is updated by the following procedure\cite{resetrev},
\begin{equation}
X_{i}(t+\Delta t)=\begin{cases}
      0 & \text{with probability $r_i \Delta t$, }\\
      X_{i}(t)+\sqrt{2 D \Delta t}\,\eta_i(t) & \text{with probability $1-r_i \Delta t$, and}~ i=1,N,\\
      \end{cases}\label{reset_wall_dyn}
\end{equation}
where $D$ is the diffusion constant which we assumed to be the same for both the boundary walls and $\eta_i(t)$ has mean and second moment, $\langle \eta_i(t)\rangle = 0$, and $\langle \eta_i(t)\eta_j(t) \rangle = \delta_{ij}$. The auto-correlation of the position $X_i(t)$ is given by\cite{resetting_correlation},
\begin{equation}
\la X_i(t)X_j(t') \ra = \delta_{ij}\frac{2 D}{r_i}e^{-r_i|t-t'|}\Big(1-e^{-r_i~\text{min}(t,t')}\Big)
\end{equation}
A stationary state is reached when $t\to \infty,~t'\to \infty$ and $|t-t'|$ is finite. In the stationary state, the auto-correlation of the position $X_i(t)$ becomes,
\bbq
\la X_i(t)X_j(t') \ra = \delta_{ij}\frac{2 D}{r_i} e^{-r_i|t-t'|}.
\ebq
The forces experienced by the boundary oscillators at the two boundaries are $f_1=kX_1$ and $f_N=kX_N$ which are stochastic hin nature. These stochastic forces are correlated as,
\bbq
\langle f_i(t) f_j(t') \rangle =\delta_{ij}\, a_i^2\, e^{-r_i|t-t'|}\quad\text{with}\quad a_i^2=\frac{2 D k^2}{r_i}\quad i=1,N. \label{eq:high_reset}
\ebq
Here $a_i$ is the strength of the noise coming from the $i$-th resetting wall. The stochastic force $f_i(t)$ acting on the boundary oscillator violates the fluctuation-dissipation relation. For a resetting driven linear chain, the equations of motion Eq.~\eqref{eq:eom} can be solved in the frequency domain,
\bbq
x_l(t)=\int_{-\infty}^{\infty} \frac{d \omega}{2 \pi} e^{-i \omega t}[G_{l1}(\omega) \tilde{f}_1(\omega)+G_{lN}(\omega) \tilde{f}_N(\omega)].
\label{eq_pos_reset}
\ebq
Here $G(\omega)$ denotes the Green's function matrix [See \ref{app_matrix} for a detailed derivation] and $\tilde{f}_i(\omega)$ is the Fourier transform of $f_i(t)$ with respect to $t$. The two-point correlation of the stochastic force $\tilde{f}_i(t)$ in the Fourier domain and can be written using Eq.~\eqref{eq:high_reset} as,
\bbq
\la \tilde{f}_i(\omega) \tilde{f}_j(\omega') \ra=2 \pi \delta_{ij}\tilde{g}(\omega,r_i)\delta(\omega+\omega')\quad\text{and}\quad \tilde{g}(\omega,r_i)=\frac{2 a_i^2 r_i}{r_i^2+\omega^2},
\ebq
where $\tilde{g}(\omega,r_i)$ is the frequency spectrum of the resetting force. 

 It can be easily shown that for the harmonic chain, the average energy current has two independent components--- a thermal component, $J_\text{therm}$ and a resetting component $J_\text{r}$. $J_\text{therm}$ is proportional to the temperature difference at the two ends of the chain, $T_1-T_N$, and $J_\text{r}$ will depend on the resetting rates of the boundary walls. The main objective of this paper is to characterize the NESS driven by the boundary resetting and we choose $T_1=T_N=0$ for the rest of the calculation. In this paper the numerical simulations are done using the scheme of \cite{lang_int} which are accurate to order $(\Delta t)^2$.

Before going into the detailed discussion, a summary of the result is presented.
\begin{itemize}
\item \textbf{Velocity fluctuation:} We measure the velocity distribution of the oscillators $P(v_l)$ in the NESS. We find that, $P(v_l)$ of the bulk oscillators approach Gaussian distribution in the thermodynamic limit, $N\to \infty$ with width $\hat{T}_\text{bulk}=m \la \dot{x}_l^2 \ra$, which is a measure of the kinetic temperature.  To characterize the finite-size dependence, we measure the kurtosis profile of velocity, $\kappa_l=\la v_l^4\ra / \la v_l^2 \ra^2-3$. We find that for the bulk oscillators, $\kappa_l$ decays as $N^{-1}$. The velocity distributions of the boundary oscillators [$P(v_1)$ or $P(v_N)$] are found to be size-independent as well as non-Gaussian. The NESS of the thermally driven chain is Gaussian, independent of the size of the harmonic chain and this is a sharp difference between the resetting driven and thermally driven case.

\item \textbf{Spatio-temporal correlation of velocity:} Another physical observable of interest is the spatio-temporal velocity correlation of the oscillators $\la v_l(t) v_{l'}(t') \ra$, in the steady state. We calculate two-time velocity correlation of single oscillator ($ \la v_l(t) v_{l}(t') \ra$) and equal-time spatial velocity correlation ($ \la v_l(t) v_{l'}(t) \ra$) for resetting driven chain explicitly. We show that in the thermodynamic limit, the two-time velocity correlation of a single oscillator in the bulk is,
\bbq
\la v_l(t) v_{l}(t') \ra=\frac{1}{ \gamma m}\int_{0}^{\pi} \frac{dq}{2 \pi} \cos \Big[\omega_c \sin{\Big(\frac{q}{2}\Big)}(t-t') \Big]\Bigg[\frac{a_1^2 r_1 }{r_1^2+\omega_c^2 \sin^2{\frac{q}{2}}}+\frac{a_N^2 r_N }{r_N^2+\omega_c^2 \sin^2{\frac{q}{2}}}\Bigg].
\ebq 
where $\omega_c=2 \sqrt{k/m}$. When $t-t'\gg \omega_c^{-1}$, $\la v_l(t) v_{l}(t') \ra \propto J_0\big(\omega_c(t-t')\big)$, here $J_0(z)$ is the Bessel function of first kind \cite{DLMF}. The equal-time spatial velocity correlation in thermodynamic limit is given by,
\begin{equation}
\la v_l(t) v_{l'}(t) \ra = \frac{1}{2 \gamma} \Big[ a_1^2 \theta(r_1,l,l')+a_N^2 \theta(r_N,l,l') \Big].
\end{equation}
where, 
\begin{equation}
\theta(r_i,l,l')=\frac{r_i}{m r_i^2+4 k}\,{}_3 \tilde{F}_2\,\Big[1/2,1,1;1-l'+l,1+l'-l;\frac{4 k}{4 k+mr_i^2} \Big].
\end{equation}
Here $l'-l$ is an integer and ${}_p \tilde{F}_q$ is a generalized regularized hypergeometric function \cite{DLMF}.

\item \textbf{Fluctuation of energy current:} We also measure the distribution of instantaneous energy current flowing through the system in the steady state. For the resetting driven oscillator chain, the distribution of instantaneous energy current at the bulk, $\mathcal{J}_l$ is found to approach
\bbq
P(\mathcal{J}_l)=\frac{1}{\pi \sqrt{ g_l}} e^{\frac{J_\text{r}}{g_l}\mathcal{J}_l}K_0\Big( \frac{u_l}{g_l}|\mathcal{J}_l|\Big).\label{eq_jdist_mt}
\ebq 
in the thermodynamic limit. Here $K_0(z)$ is the zeroth-order modified Bessel function of the second kind and the definition of $J_\text{r}$, $g_l$ and $u_l$ is given in  Eq.~\eqref{eq:curreent} and Eq.~\eqref{g_thermo}. The numerical simulation also confirms that the distribution of instantaneous energy current at the boundary $\mathcal{J}_1$, is size-independent.
\item \textbf{Effective thermal limit:} At high resetting rate, the $i$-th boundary wall acts as effective Langevin bath with effective temperature $T_i^\text{eff}$. In this limit, the known result of the average energy current for a thermally driven oscillator chain is recovered \cite{rllmothRieder}. The equal-time spatial velocity correlation also becomes uncorrelated for $l \neq l'$ in this effective thermal limit.

\end{itemize}

\section{Velocity distribution and kurtosis profile of the velocity}\label{sec_vel_dist}

The probability distributions of the velocity of the constituent oscillators play an important role in characterizing the NESS of the driven oscillator chain. For thermally or activity-driven oscillator chains, the velocity distributions of the oscillators at the bulk are found to be Gaussian with its width given by local kinetic temperature $\hat{T}_\text{bulk}=m\la v_l^2 \ra$ \cite{rllmothRieder,activity_stationary}. However, the velocity distribution of boundary oscillators depends on the specific dynamics of driving. For resetting driven harmonic chain also, we first measure the velocity distributions of bulk and boundary oscillators for different system sizes and different resetting rates at the two boundaries.
\subsection{Distribution of velocity}\label{vel_dist}

\begin{figure}[ht]
\centering
    \includegraphics[width=0.85\linewidth]{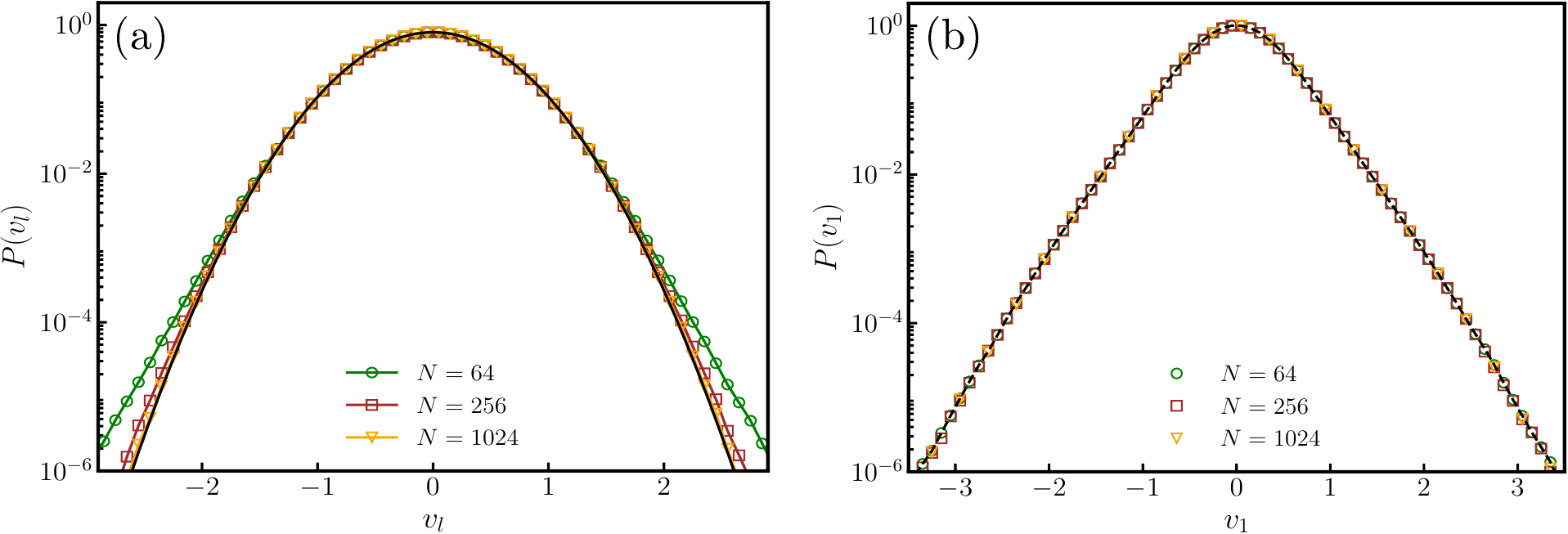} 

    \caption{ (a) Velocity distribution of the bulk oscillator, $l=N/2$ measured from simulation with chains of different sizes for the same resetting rate, $r_1=r_N=2.5$.  The black solid line corresponds to the Gaussian distribution for thermodynamically large system size ($N\to \infty$) given by Eq.~\eqref{eq_Gauss}. (b) Velocity distribution of the boundary oscillator, $l=1$ measured from simulation with chains of different sizes for the same resetting rate, $r_1=r_N=2.5$.  Other parameters are $D=m=\gamma=k=1$.}
\label{fig:vel_dist}
\end{figure}

\begin{figure}[ht]
\centering
    \includegraphics[width=0.425\linewidth]{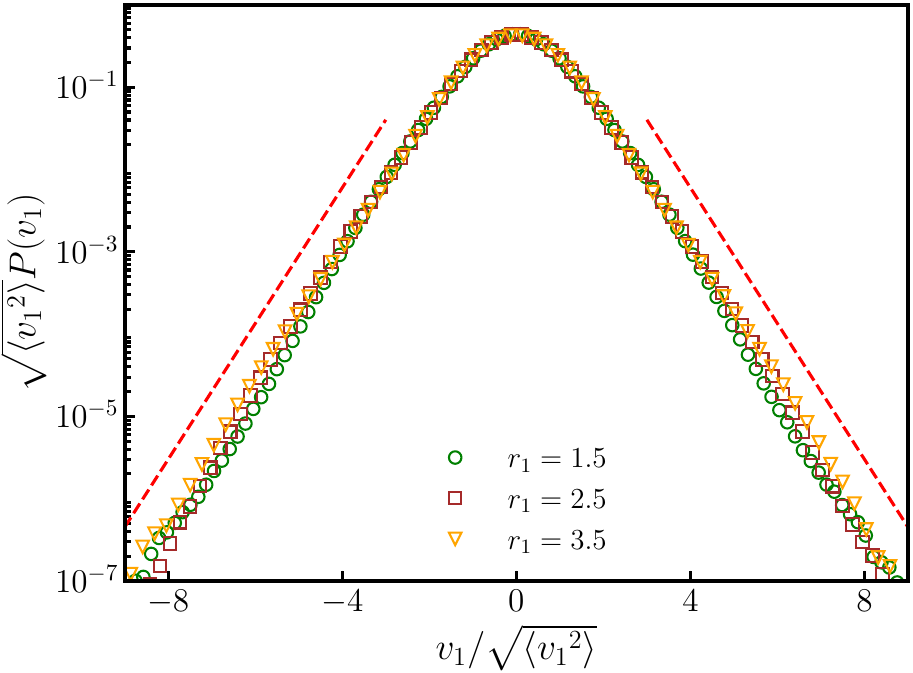} 

    \caption{ The plot of the scaled velocity distribution of the oscillator at the left boundary for fixed $r_N=2.5$ and three different values of $r_1$. The symbols denote the numerical simulation with a chain of $N=1024$. Other parameters are $D=m=\gamma=k=1$.}
\label{fig:vel_dist_boundary}
\end{figure}

The velocity of the $l$-th oscillator for resetting driven chain is [from Eq.~\eqref{eq_pos_reset}],
\bbq
v_l(t)=\int_{-\infty}^{\infty} \frac{d \omega}{2 \pi}e^{-i\omega t}(-i \omega)\big[G_{l1}(\omega)\tilde{f}_1(\omega)+G_{lN}(\omega)\tilde{f}_N(\omega)\big],\label{eq_velocity}
\ebq
and we numerically measure the velocity distribution of the bulk oscillators $P(v_l)$ for different system sizes. In Fig.~\ref{fig:vel_dist}(a), the numerically measured $P(v_l)$ is plotted for different system sizes. For any finite $N$, $P(v_l)$ shows significant deviation at the tails from the Gaussian distribution with standard deviation $\hat{T}_\text{bulk}$. However, the deviation tends to decrease with an increase in system size. Therefore we conclude that,
\bbq
\text{For}~N\to \infty,~~P(v_l)=\frac{1}{\sqrt{2 \pi \hat{T}_{bulk}}}\exp{\Bigg(\frac{-m v_l^2}{2 \hat{T}_\text{bulk}}\Bigg)}\label{eq_Gauss},
\ebq
where the value of kinetic temperature at the bulk is [see \ref{appendix_two_time_correl_reset} for the derivation],
\bbq
\hat{T}_\text{bulk}=\frac{a_1^2 }{2 \gamma\sqrt{r_1^2+\frac{4 k }{m}}}+\frac{a_N^2 }{2 \gamma\sqrt{r_N^2+\frac{4 k }{m}}}.\label{eq_bulk_kin_temp}
\ebq
This finite-size dependence of velocity distribution makes the resetting driven harmonic chain different than the thermally driven harmonic chain, where the steady state is Gaussian irrespective of the size. Numerical simulations with different system sizes suggest, unlike velocity distributions of the bulk oscillators, $P(v_l)$, the velocity distribution of the boundary oscillator $P(v_1)$ [or $P(v_N)$] is size-independent, [see  Fig.~\ref{fig:vel_dist_boundary}].

The scaled velocity distribution of the oscillator at the left boundary is shown in Fig.~\ref{fig:vel_dist_boundary}. Here the numerically measured velocity distribution $P(v_1)$ is scaled with numerically measured $\sqrt{\la v_1^2 \ra}$, which shows a scaling collapse. The numerical simulation also indicates that the velocity distribution of the boundary oscillator has an exponentially decaying tail [which is shown with the red dashed line in Fig.~\ref{fig:vel_dist_boundary}]. 

\subsection{Kurtosis profile of velocity}\label{kurt}

\begin{figure}[ht]
\centering
    \includegraphics[width=0.85\linewidth]{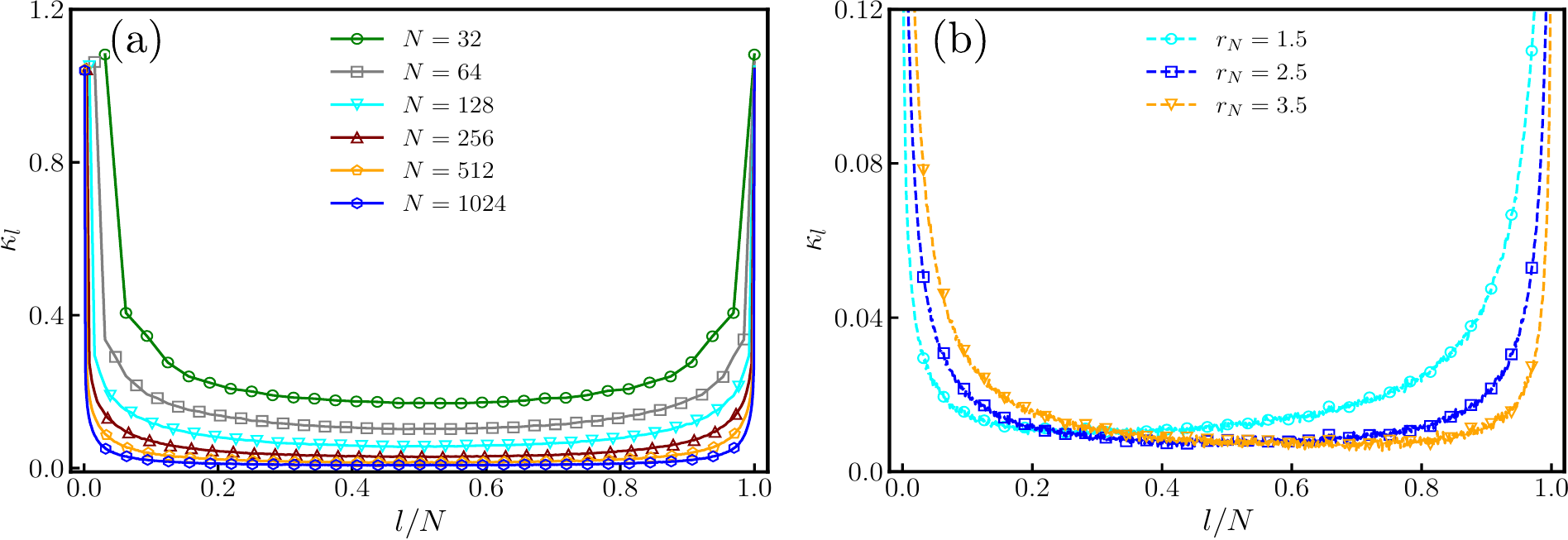} 

    \caption{(a) Numerically measured velocity kurtosis profile for the same resetting rate at the two ends, $r_1=r_N=2.5$ with chains of different sizes. (b) Velocity kurtosis profile for a chain of $N=1024$ for fixed $r_1=2.5$ and different $r_N$, other parameters are $D=m=\gamma=k=1$. }
\label{fig:kurt_prof}
\end{figure}

To characterize the finite-size dependence of the velocity distribution of the bulk oscillators, we measure the kurtosis profile of velocity. The definition of the kurtosis of the velocity of the $l$-th oscillator is,
\bbq
\kappa_l=\frac{\la v_l^4 \ra}{\la v_l^2 \ra^2}-3.
\ebq 
If the velocity distribution of the $l$-th oscillator is Gaussian, then the $\kappa_l=0$. Typically the distribution will have a fatter tail when $\kappa_l>0$. We compute the velocity kurtosis profile for different system sizes using numerical simulation. In Fig.~\ref{fig:kurt_prof}(a), $\kappa_l$ for different system sizes is plotted for $r_1=r_N$ and from this plot, it is clear that the value of the kurtosis at the bulk decreases as the system size increases. In Fig.~\ref{fig:kurt_prof}(b) $\kappa_l$ for a chain of length $N=1024$ is plotted for a fixed value of $r_1$ and three different values of $r_N$. Note that, the velocity kurtosis profile is symmetric for $r_1=r_N$ and the velocity kurtosis profile is symmetric in $|r_1-r_N|$. It is clear from Fig.~\ref{fig:kurt_prof}(b) that the velocity distribution of the bulk oscillators will reach the Gaussian distribution if the resetting rate is high. However, the boundary oscillators will have a fatter tail even at a higher resetting rate.
\begin{figure}[ht]
\centering
    \includegraphics[width=0.425\linewidth]{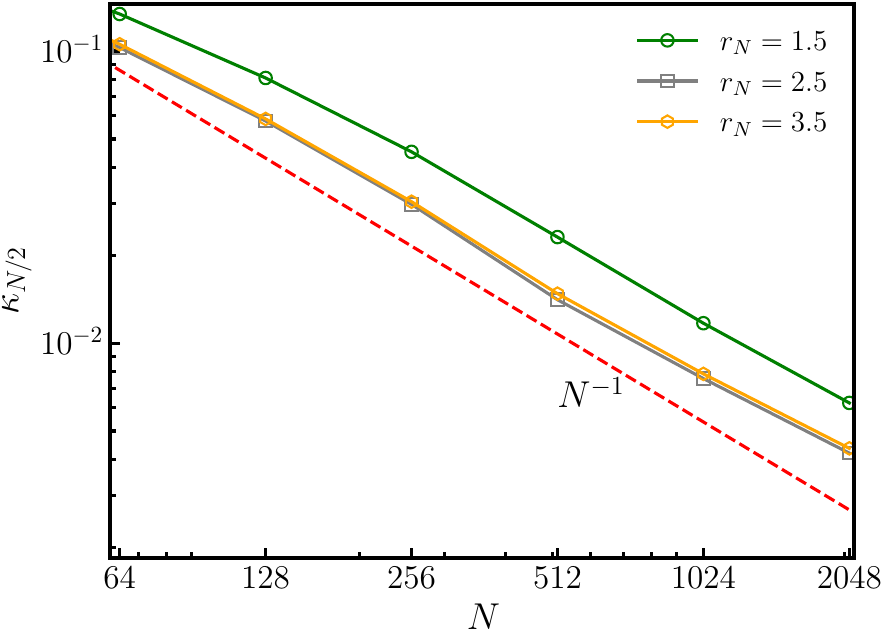} 

    \caption{Numerically measured velocity kurtosis of the $l$-th oscillator at the bulk (where $l=N/2$) as a function of system size $N$ for fixed $r_1=2.5$ and different value of $r_N$. Other parameters are $D=m=\gamma=k=1$ and the red dashed line shows $\sim N^{-1}$ decay.}
\label{fig:kurt_decay}
\end{figure}

It is intriguing to investigate the system size dependence of the velocity kurtosis profile, therefore we numerically estimate $\kappa_\text{bulk}$ [$\kappa_l$ at $l=N/2$] with increasing system size $N$ for fixed $r_1$ and three different values of $r_N$. The result is shown in Fig.~\ref{fig:kurt_decay}. From the result of numerical simulation, we conclude that,
\bbq
\kappa_{\text{bulk}}\sim N^{-1}.
\ebq
Therefore only in the thermodynamic limit, the velocity distributions of the bulk oscillators become Gaussian.  Size-dependent NESS is the signature of the resetting driven harmonic chain because such behavior is absent for the thermally driven case.

\section{Spatio-temporal correlation of velocity}\label{sec_cross}

Correlation functions play an important role in characterizing transport processes. For example, the integral of equilibrium velocity-velocity correlation $\la v(t)v(t') \ra$ is related to the diffusion coefficient of a Brownian particle. For thermally driven transport in the mass-disordered harmonic chain, the two-time correlation of the current determines the asymptotic size dependence of the current fluctuation in the steady state \cite{time_correl}. In this section, we discuss the two-point spatio-temporal correlation of velocity $ \la v_l(t) v_{l'}(t') \ra$ in the bulk. 

Using Eq.~\eqref{eq_velocity}, we can write the two-point spatio-temporal correlation of velocity as,
\bbq
\la v_l(t)v_{l'}(t') \ra = \int_{-\infty}^{\infty} \frac{d\omega}{2 \pi} \omega^2 e^{-i\omega (t-t')}\big[G_{l1}(\omega)G_{l'1}^*(\omega)\tilde{g}(\omega,r_1)+G_{lN}(\omega)G_{l'N}^*(\omega)\tilde{g}(\omega,r_N)\big].\label{eq_cross_v_reset_mt}
\ebq
In the following, we calculate the two-time velocity correlation of a single oscillator $ \la v_l(t) v_{l}(t') \ra$ and equal-time spatial velocity correlation $ \la v_l(t) v_{l'}(t) \ra$ in the stationary state explicitly for the resetting driven chain.

\subsection{Two-time velocity correlation of single oscillator}
In the stationary state, the two-time velocity correlation of a single oscillator, $\la v_l(t) v_l(t') \ra$ is a function of $t-t'$ [see Eq.~\eqref{eq_cross_v_reset_mt}]. To simplify the calculation, we consider $\la v_l(t) v_{l}(0) \ra$ which can be written as,
\begin{equation}
\la v_l(t) v_{l}(0) \ra=\int_{-\infty}^{\infty} \frac{d\omega}{2 \pi} \omega^2 e^{-i\omega t}\big[|G_{l1}(\omega)|^2\tilde{g}(\omega,r_1) +|G_{lN}(\omega)|^2 \tilde{g}(\omega,r_N)\big]. \label{cd_base}
\end{equation}
\begin{figure}[ht]
\centering
    \includegraphics[width=0.85\linewidth]{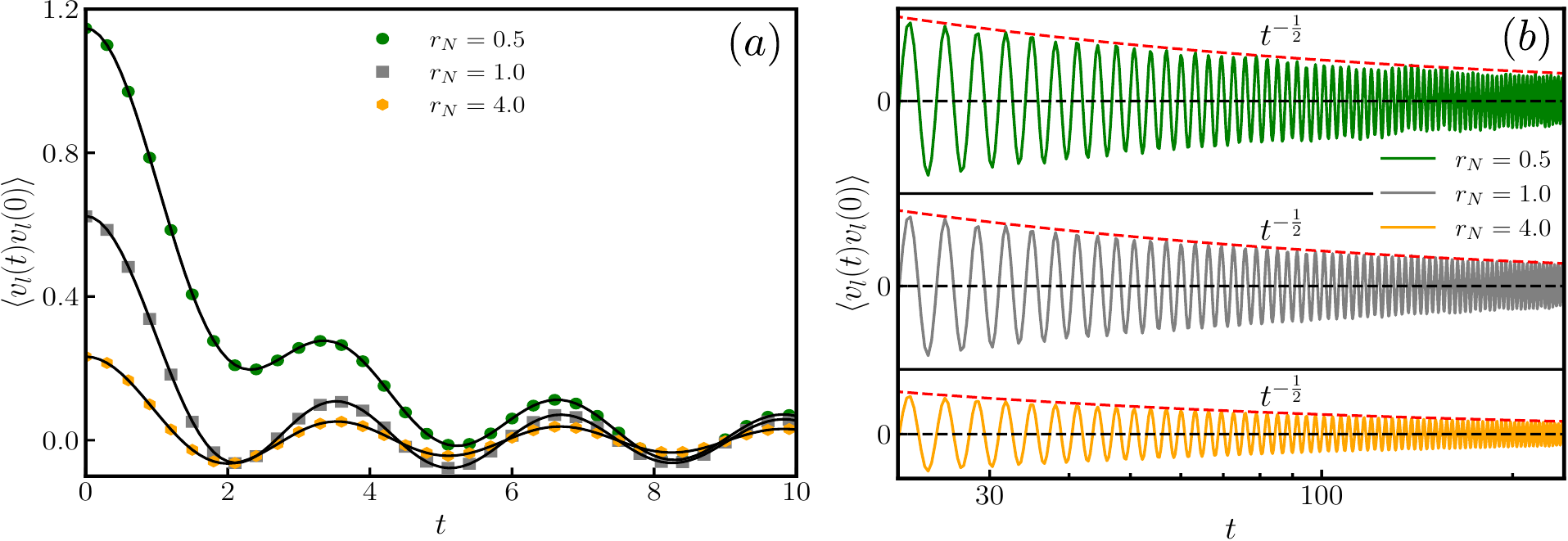} 

    \caption{(a) Plot of $\la v_l(t) v_{l}(0) \ra$ as a function of time for fixed $r_1=1.0$ and three different value of $r_N$. Symbols correspond to the data obtained from numerical simulation with a chain of $N=256$. The black solid line corresponds to Eq.~\eqref{eq_reset_Cd}. (b) Plot of $\la v_l(t) v_{l}(0) \ra$ in the large time limit is shown along with the red dashed line which corresponds to $1/\sqrt{t}$ behavior. The other parameters are $D=m=\gamma=k=1$.}
\label{fig:correl}
\end{figure}
The above equation has two separate contributions coming from two resetting walls at the two boundaries and we can write,
\bbq
\la v_l(t) v_{l}(0) \ra= a_1^2{C}(r_1,t)+a_N^2{C}(r_N,t).
\ebq
where ${C}(r_i,t)$ is the contribution from the $i$-th resetting wall.  For the oscillators in the bulk [$1\ll l\ll N$], $C(r_i,t)$ can be computed explicitly and takes a simple form in the thermodynamic limit,
\bbq
{C}(r_i,t) =\frac{r_i}{ \gamma m}\int_{0}^{\pi} \frac{dq}{2 \pi} \frac{\cos \Big[\omega_c t \sin{\Big(\frac{q}{2}\Big)} \Big] }{r_i^2+\omega_c^2 \sin^2{\frac{q}{2}}},
\ebq
see \ref{app_reset} for the detailed derivation. Combining the contributions from both the reservoirs, we can write $\la v_l(t) v_{l}(0) \ra$ in the thermodynamic limit as,
\bbq
\la v_l(t) v_{l}(0) \ra=\frac{1}{ \gamma m}\int_{0}^{\pi} \frac{dq}{2 \pi} \cos \Big[\omega_c t \sin{\Big(\frac{q}{2}\Big)} \Big]\Bigg[\frac{a_1^2 r_1 }{r_1^2+\omega_c^2 \sin^2{\frac{q}{2}}}+\frac{a_N^2 r_N }{r_N^2+\omega_c^2 \sin^2{\frac{q}{2}}}\Bigg]\label{eq_reset_Cd}.
\ebq
The above integral can be evaluated numerically for arbitrary values of $t$. This is illustrated in Fig~\ref{fig:correl}(a) where the symbols correspond to the data obtained from numerical simulation for a fixed $r_1$ and different values of $r_N$. The black solid lines in Fig~\ref{fig:correl}(a) correspond to the numerical integration of Eq.~\eqref{eq_reset_Cd}. 

It is difficult to get the closed form of Eq.~\eqref{eq_reset_Cd}. However, it is possible to predict the small and large time behavior of $ \la v_l(t) v_{l}(0) \ra$. For very small value of $t$ or $t\ll \omega^{-1}_c$, $\cos [\omega_c t \sin{(q/2)} ]$ can be expanded up to second order in time $t$. Therefore, for small value of $t$, $\la v_l(t) v_{l}(0)$ is proportional to $t^2$. Similarly, when $t\gg \omega_c^{-1}$, the integral can be performed exactly [see \ref{appendix_two_time_correl_reset} for the detailed derivation]. In this large time limit,
\bbq
\la v_l(t) v_{l}(0) \ra \simeq \Big[\frac{a_1^2 r_1}{2 \gamma (m r_1^2+4 k)}+ \frac{a_N^2 r_N}{2 \gamma (m r_N^2+4 k)} \Big]J_0(\omega_c t),
\ebq 
where $\omega_c=2\sqrt{k/m}$, and $J_0(z)$ is the Bessel function of the first kind. In Fig~\ref{fig:correl}(b) numerically measured $\la v_l(t) v_{l}(0) \ra$ ($l=N/2$) is plotted for large $t$, which is an oscillatory function of $t$ and the envelope of $\la v_l(t) v_{l}(0) \ra$ seemed to decay as $t^{-\frac{1}{2}}$, this is similar to the behavior of $J_0(\omega_c t)$ when $t\gg \omega_c^{-1}$.

\subsection{Equal-time spatial velocity correlation}
\begin{figure}[ht]
\centering
    \includegraphics[width=0.425\linewidth]{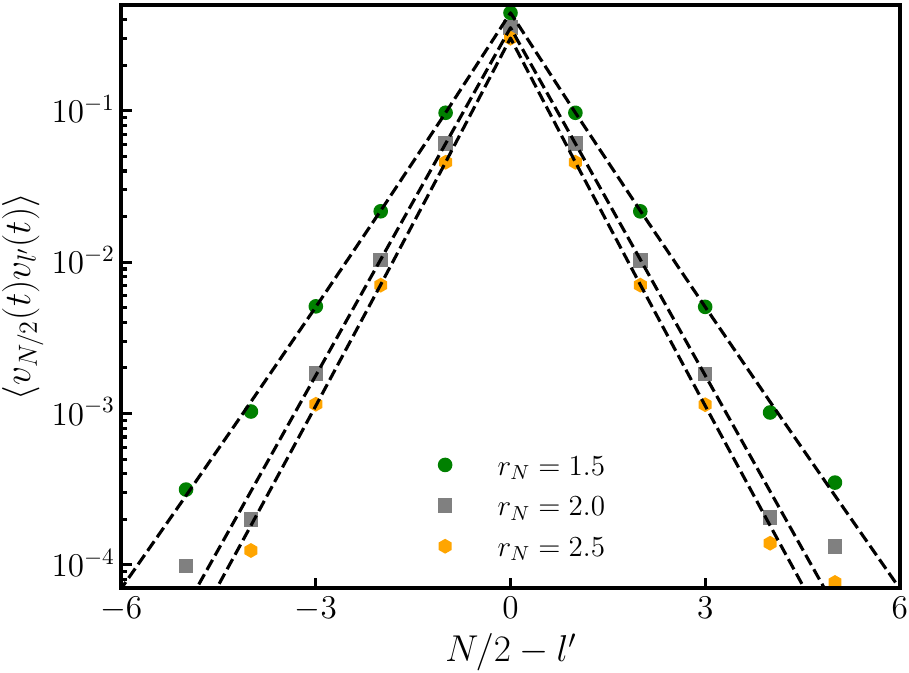} 

    \caption{Plot of $\la v_{N/2}(t) v_{l'}(t) \ra$ as a function of $N/2-l'$ for fixed $r_1=2.0$ and three different value of $r_N$. Symbols correspond to the data obtained from numerical simulation with a chain of size $N=64$. The black dashed lines correspond to Eq.~\eqref{eq_reset_Cs} with $l-l'$ as an integer. The other parameters are the same as the previously mentioned simulations.}
\label{fig:correl2}
\end{figure}
Equal-time spatial velocity correlation, $\la v_l(t)v_{l'}(t) \ra$, can be calculated by taking $t=t'$ in Eq.~\eqref{eq_cross_v_reset_mt}.
\begin{equation}
\la v_l(t)v_{l'}(t) \ra= \int_{-\infty}^{\infty} \frac{d\omega}{2 \pi} \omega^2 [G_{l1}(\omega)G_{l'1}^*(\omega)\tilde{g}(\omega,r_1)+G_{lN}(\omega)G_{l'N}^*(\omega)\tilde{g}(\omega,r_N)].\label{cs_base}
\end{equation} 
In the stationary state $\la v_l(t)v_{l'}(t) \ra$ does not depend on the time and has two separate contributions coming from the two walls performing intermittent resetting. Therefore $\la v_l(t)v_{l'}(t) \ra $ can be written as, 
\bbq
\la v_l(t)v_{l'}(t) \ra= a_1^2\theta(r_1,l,l')+a_N^2\theta(r_N,l,l').\label{limit_Cs2}
\ebq
In the bulk, i.e. $1\ll l ,l' \ll N$, the contribution from the $i$-th wall in the limit $N \rightarrow \infty$ is [see \ref{app_reset} for detailed derivation],
\begin{equation}
\theta(r_i,l,l')=\frac{r_i^2}{2 \pi \gamma}\int_{0}^{\pi} dq \frac{\cos{(l'q-lq)}}{m r_i^2+4 k \sin^2\frac{q}{2}}\label{limit_Cs}
\end{equation}
where, 
\begin{equation}
\theta(r_i,l,l')=\frac{r_i}{m r_i^2+4 k}\,{}_3 \tilde{F}_2\,\Big[1/2,1,1;1-l'+l,1+l'-l;\frac{4 k}{4 k+mr_i^2} \Big]~\text{and}~ l'-l\in \mathbb{I}.
\end{equation}

Here ${}_p \tilde{F}_q$ is a generalized regularized hypergeometric function \cite{DLMF}. Combining the contribution of both reservoirs we arrive at the,
\begin{equation}
\la v_l(t) v_{l'}(t) \ra =  \frac{a_1^2}{2 \gamma} \theta(r_1,l,l')+\frac{a_N^2}{2 \gamma} \theta(r_N,l,l') .\label{eq_reset_Cs}
\end{equation}
In Fig~\ref{fig:correl2} we plotted numerically measured $\la v_l(t) v_{l'}(t) \ra$ for a fixed value of $r_1$ and different values of $r_N$ which matches with the analytic prediction,  Eq.~\eqref{eq_reset_Cs} well. 

The two time velocity correlation of single oscillator $\la v_l(t) v_{l}(0) \ra$ and the equal-time spatial velocity correlation $\la v_l(t) v_{l'}(t) \ra$ in the stationary state depends on the frequency spectrum of the nonequilibrium reservoir $\tilde{g}(\omega,r_i)$ [Eq.~\eqref{eq_cross_v_reset_mt}]. Therefore, these two observables can be evaluated for the nonequilibrium driving described in Ref.~\cite{activity_driven_chain,activity_stationary}.

\section{Fluctuation and second moment of instantaneous energy current}\label{sec_current}
The definition of instantaneous energy current from the left reservoir to the system is,
\bbq
\mathcal{J}_{1}=(-\gamma v_{1} + f_{1})v_{1}.\label{boundary_current}
\ebq
\begin{figure}[ht]
    \centering
    \includegraphics[width=0.425\linewidth]{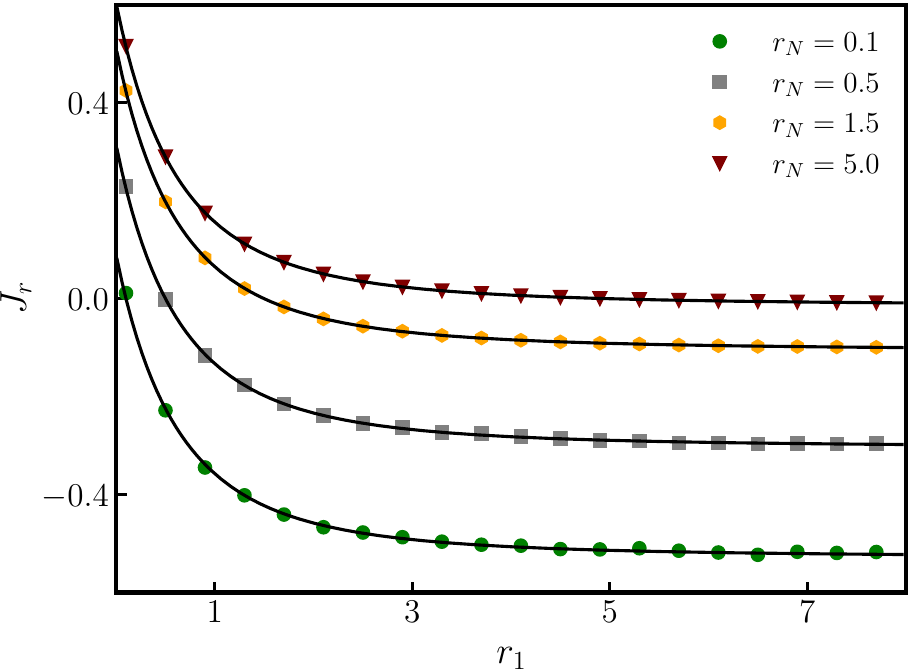}
    \caption{Plot of average energy current $J_\text{r}$ as functions of resetting rate $r_1$ for different values of $r_N$. Symbols obtained from the numerical simulation with $N=1024$ oscillators and other parameters are $D=m=\gamma=k=1$. Black solid lines correspond to Eq.~\eqref{eq:curreent}.}
    \label{fig:current_sweep}
\end{figure}
Similarly instantaneous energy current from the right reservoir to the system is
\bbq
\mathcal{J}_{N+1}=(-\gamma v_{N}+ f_{N})v_{N}.
\ebq
In the bulk, the expression for the instantaneous energy current between $l-1$th and $l$-th oscillator,
\bbq
\mathcal{J}_{l}=\frac{k}{2}\left(v_{l-1}+v_{l}\right) \left(x_{l-1}-x_{l} \right)\label{bulk_current}.
\ebq
The steady state of the resetting-driven harmonic chain is characterized by the existence of non-zero average energy flux through the chain. In NESS, the average energy current flowing through the chain is,
\bbq
J_\text{r}=\la \mathcal{J}_{1}\ra = \la \mathcal{J}_{2} \ra =\cdots =\la \mathcal{J}_{l} \ra = \cdots = -\la \mathcal{J}_{N+1} \ra. \label{eq_avg_current_sys}
\ebq
The analytical expression of the average energy current $J_\text{r}$ can be calculated easily following the method described in Ref.~\cite{activity_driven_chain}. For completeness, we quote the result here.\begin{eqnarray}
J_\text{r}&=&\frac{m}{4 \gamma ^2} \Big[a_1^2{\zeta}(r_1)-a^2_{N}{\zeta}(r_N)\Big]\quad\text{where}\cr
{\zeta}(r_j)&=&\frac{r_j k^2 \left( \sqrt{1+\frac{4\gamma^2}{mk} }-1 \right)+r_j^3\gamma^2\left( 1-\sqrt{1+\frac{4 k}{mr_j^2} } \right)}{(k^2-r_j^2\gamma^2)}\label{eq:curreent}.
\end{eqnarray}

In Fig.\ref{fig:current_sweep}, we have shown the plot of average energy current $J_\text{r}$ for fixed $r_1$ and different $r_N$ along with the analytical prediction Eq.~\eqref{eq:curreent}. In the next subsection, we will discuss the distribution of the instantaneous energy current at the bulk and the boundary.

\subsection{Distribution of instantaneous energy current at the bulk}\label{subsec_bulk_current_dist}

In this section, we will concentrate on the stationary state distribution of instantaneous current at the bulk $P(\mathcal{J}_{l})$. It is clear from the definition of $\mathcal{J}_{l}$ mentioned in Eq.~\eqref{bulk_current} that the $P(\mathcal{J}_{l})$ will depend on the joint distribution  $\mathcal{P}(\{x_l,v_l\})$. In Sec.~\ref{sec_vel_dist}, we have shown that the velocity distribution of the bulk oscillators of the resetting driven harmonic chain approaches the Gaussian distribution Eq.~\eqref{eq_Gauss} for thermodynamically large system size. A similar result is also expected for the position variables $x_l$ in the bulk. Therefore, we assume that the joint probability distribution of $\{x_l,v_l\}$ of the bulk oscillators, $\mathcal{P}(\{x_l,v_l\})$ is a multivariate Gaussian in the thermodynamic limit. Therefore, we can write the joint distribution of $\{v_{l-1},v_{l},x_{l-1},x_{l}\}$ in that limit as,
\bbq
\mathcal{P}(v_{l-1},v_{l},x_{l-1},x_{l})=\frac{\exp\left[ -\frac{1}{2} {W}_l^T\mathcal{E}^{-1}_l{W}_l  \right]}{\sqrt{(2 \pi)^4 \text{det}(\mathcal{E}_{l})}}, \label{eq: joint_dist_bulk}
\ebq
where, $W_l^T=(v_{l-1}~v_{l}~x_{l-1}~x_{l})$ and $\mathcal{E}_{l}=\la W_l W_l^T \ra$, the corresponding $4\times 4$ correlation matrix. For Gaussian NESS, the instantaneous current distribution $P(\mathcal{J}_l)$ at the bulk is derived in Ref.~\cite{activity_stationary}. Here we quote the final result,
\bbq
P(\mathcal{J}_l)=\frac{1}{\pi\sqrt{ g_l}} \exp\Big({\frac{J_\text{r}}{g_l}\mathcal{J}_l\Big)}K_0\Big( \frac{u_l}{g_l}|\mathcal{J}_l|\Big)\label{eq_bulk_j_dist}.
\ebq 
Here $K_0(z)$ is the zeroth-order modified Bessel function of
the second kind, and $J_\text{r}$ is the average energy current flowing through the resetting driven harmonic chain. Note that, the specific form of distribution function mentioned in Eq.~\eqref{eq_bulk_j_dist} also appeared in the context of time-integrated heat current fluctuations of Brownian particles in an active environment \cite{heat_dist}. For resetting driven chain, the $g_l$ and $u_l$ appeared in Eq.~\eqref{eq_bulk_j_dist} are given by,
\bbq
g_l=\frac{k \hat{T}_{\text{bulk}}}{2}[\hat{T}_\text{bulk}+\la v_{l-1} v_{l} \ra]-J_\text{r}^2,~~\text{and}~~u_{l}=\sqrt{ g_l+J_\text{r}^2}.\label{g_thermo}
\ebq
Here $\la v_{l-1} v_{l} \ra$ is equal time spatial velocity correlation for a driven harmonic chain.

\begin{figure}[ht]
    \centering
    \includegraphics[width=0.85\linewidth]{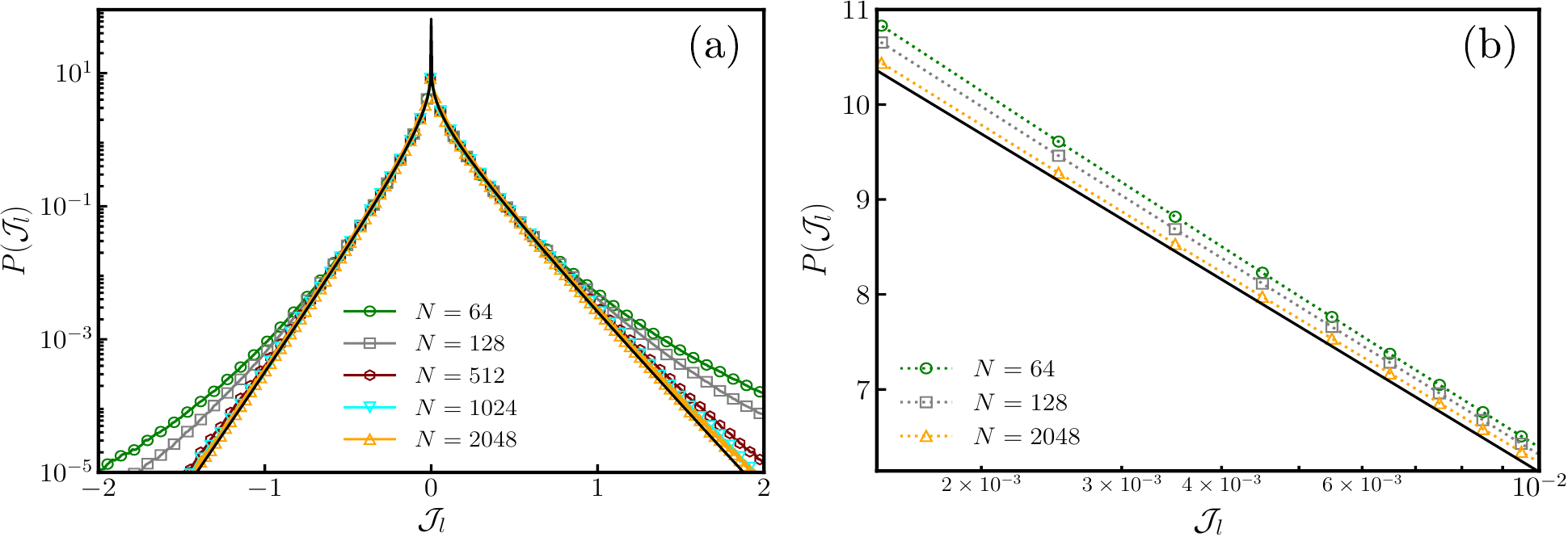}
    \caption{(a) Numerically measured distribution of instantaneous current distribution at bulk $P(\mathcal{J}_l)$ for different resetting rate at boundary, $r_1=2.5$ and $r_N=3.5$. In (b), the behavior of $P(\mathcal{J}_l)$ near the origin is shown. The black solid line in both (a) and (b) corresponds to Eq.~\eqref{eq_bulk_j_dist} and Eq.\eqref{jdist_origin} respectively.  Other parameters are set at $D=m=\gamma=k=1$.}
    \label{fig:current_dist}
\end{figure}

In Fig.~\ref{fig:current_dist}(a), numerically measured $P(\mathcal{J}_l)$ of different system sizes are plotted for different values of $r_1$ and $r_N$ and for a large system size (e.g. $N=1024$), there is a good agreement with Eq.~\eqref{eq_bulk_j_dist}, shown by the black solid line.  This serves as a cross-check of our assumption that in the thermodynamic limit, the joint distribution $\mathcal{P}(\{v_l,x_l\})$ for the resetting driven harmonic chain, is a multivariate Gaussian, where $1\ll l\ll N$. It is noteworthy that such strong finite-size dependence of $P(\mathcal{J}_l)$ is not reported for activity-driven harmonic chain \cite{activity_stationary}. This finite-size dependence is absent for thermally driven harmonic chains as the NESS is Gaussian irrespective of the size of the system. For small system sizes, deviations from the analytic curve are visible on the tails of the distributions which decreases with the increase of the system size.  

Using the asymptotic behavior of $K_0(z)$ for $z\to 0$, we get,
\bbq
P(\mathcal{J}_l)=-\frac{1}{\pi\sqrt{ g_l}}\left(\ln{\frac{u_l}{2 g_l}|\mathcal{J}_l|}+E_{\gamma} \right)+O(\mathcal{J}_l)\label{jdist_origin},
\ebq
near $\mathcal{J}_l=0$. Here $E_{\gamma} \simeq 0.577216 $ is the Euler's constant \cite{DLMF}. From this figure, it is clear that $P(\mathcal{J}_l)$ has a logarithmic divergence even for small system sizes and approaches Eq.~\eqref{jdist_origin} with increasing system size.

\subsection{Distribution of the instantaneous energy current at the boundary}
The definition of instantaneous current from the left reservoir to the system is, 
\bbq
\mathcal{J}_{1}=(-\gamma v_{1}+ f_{1})v_{1},
\ebq
and corresponding distribution can be written as, 
\bbq
P(\mathcal{J}_1)=\int dv_1 df_1 \, \delta [\mathcal{J}_1-(-\gamma v_1+f_1) v_1]\,\mathcal{P}(v_1,f_1). \label{eq: j_dist_bound}
\ebq
We mentioned in Sec.~\ref{vel_dist} that the velocity distributions of the boundary oscillators [$P(v_1)$ or $P(v_N)$] are independent of the size of the system. It is important  to note that, in the stationary state, the position distribution $P(X_i)$ of the boundary wall [whose dynamics is given in Eq.~\eqref{reset_wall_dyn}] is given by \cite{satyaprl_reset},
\bbq
P(X_i)=\sqrt{\frac{r_i}{4 D}}\exp{\Big(-\sqrt{\frac{r_i}{D}}|X_i|\Big)}\label{eq_reset_dist}.
\ebq
Therefore the force $f_1=k X_1$ and $f_N=k X_N$ exerted on the boundary oscillators are of the same form and independent of the size of the oscillator chain. 

\begin{figure}[ht]
    \centering
    \includegraphics[width=0.85\linewidth]{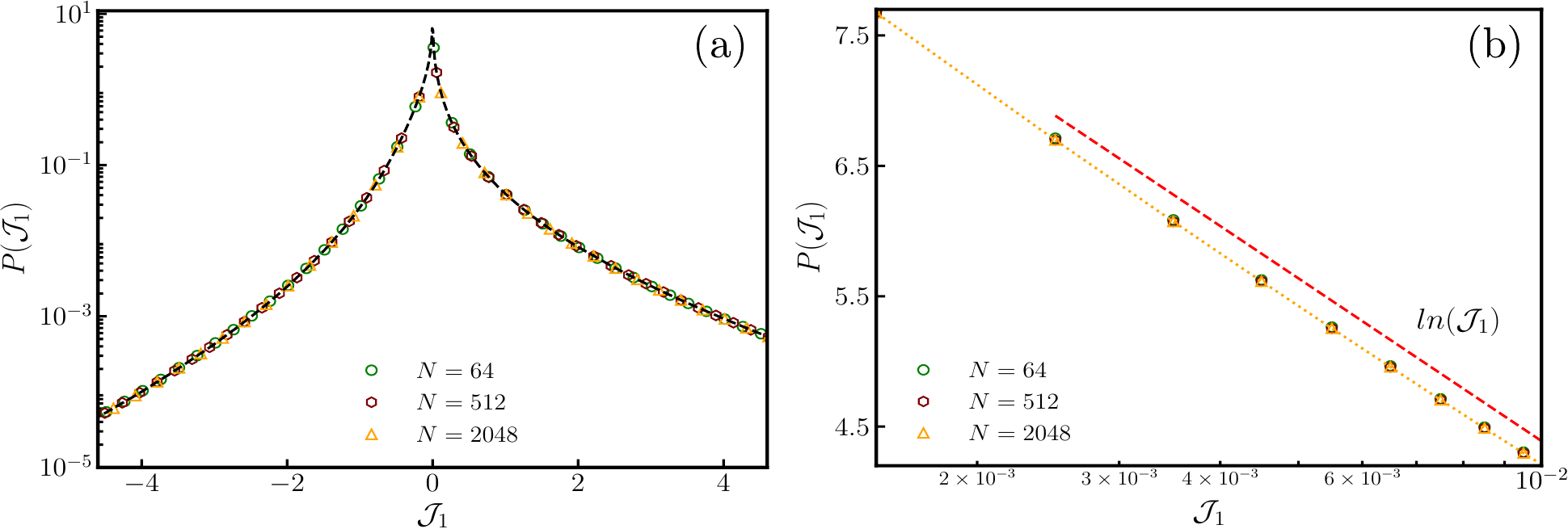}
    \caption{(a) Numerically measured distribution of instantaneous current distribution at left boundary $P(\mathcal{J}_1)$ for $r_1=2.5$ and $r_N=3.5$. (b) Behavior of $P(\mathcal{J}_1)$ near the origin. The other parameters are set at $D=m=\gamma=k=1$.}
    \label{fig:current_dist_boundary}
\end{figure}
As the marginal distributions $P(v_1)$ and $P(f_1)$ are both size-independent, it can be inferred that the joint distribution $\mathcal{P}(v_1,f_1)$ is size-independent. Therefore, the distribution of instantaneous current at the boundary will also be size-independent [see Eq.~\eqref{eq: j_dist_bound}]. The analytical form of $P(\mathcal{J}_1)$ is hard to calculate as the analytical form of $P(v_1)$ is unknown. However, in Fig.~\ref{fig:current_dist_boundary}(a), the numerical simulation result for $P(\mathcal{J}_1)$ with different system size is shown, which confirms that the $P(\mathcal{J}_1)$ is size-independent. Behavior of $P(\mathcal{J}_1)$ near origin is shown in Fig.~\ref{fig:current_dist_boundary}(b). Similar to $P(\mathcal{J}_l)$, $P(\mathcal{J}_1)$ also has logarithmic divergence near the origin.

\subsection{Second moment of the energy current}

The second moment of the instantaneous energy current at bulk $\mathcal{J}_l$ and boundary $\mathcal{J}_1$ can be calculated from the moment generating functions given by,
\bbq
\la \mathcal{J}_l^2 \ra =-\frac{d^2}{d \mu^2} \la e^{i \mu \mathcal{J}_l} \ra \Big|_{\mu=0},
\ebq
\bbq
\la \mathcal{J}_1^2 \ra =-\frac{d^2}{d \mu^2} \la e^{i \mu \mathcal{J}_1} \ra \Big|_{\mu=0}.
\ebq
\begin{figure}[ht]
    \centering
    \includegraphics[width=0.85\linewidth]{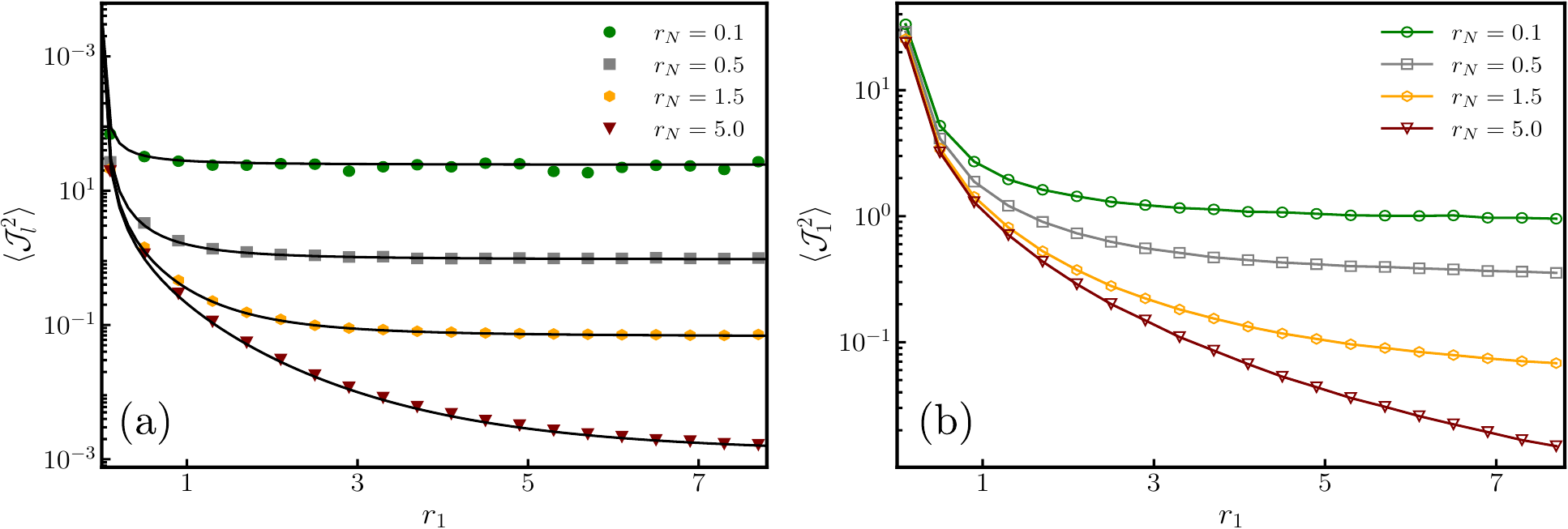}
    \caption{(a) Plot of second moment of $\mathcal{J}_l$ as functions of resetting rate $r_1$. The black solid line corresponds to the Eq.~\eqref{eq_bulk_moment}. (b) The plot of the second moment of $\mathcal{J}_1$ as functions of resetting rate $r_1$. Symbols obtained from the numerical simulation with $N=1024$ oscillator and for all numerical simulations other parameters are $D=m=\gamma=k=1$. }
    \label{fig:j2_sweep}
\end{figure}
In Sec.~\ref{subsec_bulk_current_dist}, it has been shown that the current distribution $P(\mathcal{J}_l)$ approaches Eq.~\eqref{eq_bulk_j_dist} for thermodynamically large system and the second moment of $\mathcal{J}_l$ is given by \cite{activity_stationary}, 
\bbq
\la \mathcal{J}_l^2 \ra =2 J_\text{r}+u_l^2. \label{eq_bulk_moment}
\ebq
In Fig.~\ref{fig:j2_sweep}(a), numerically measured $\la \mathcal{J}_l^2 \ra$ is shown as a function of $r_1$ for different values of $r_N$ along with the analytical plot of Eq.~\eqref{eq_bulk_moment}, and they are in a good agreement.

We do not have the analytical expression of the second moment of $\mathcal{J}_1$. However, $\la \mathcal{J}_1^2\ra$ is size-independent as the probability distribution $P(\mathcal{J}_1)$ is size-independent. In Fig.~\ref{fig:j2_sweep}(b) numerically measured $\la \mathcal{J}_1^2 \ra$ is shown as a function of $r_1$ for different values of $r_N$.

\section{Effective thermal limit}\label{sec_thermal_case}

In the limit of a high resetting rate, the colored noise coming from the resetting wall at the boundary becomes an effective white noise. Therefore an effective fluctuation-dissipation relation can be written in terms of an effective temperature, $T_i^\text{eff}$.
\bbq
\langle f_i(t) f_j(t') \rangle =\delta_{ij} a_i^2 e^{-r_i|t-t'|}\xrightarrow{r_i\to \infty} 2 \gamma \delta_{ij} T_i^\text{eff} \delta(t-t')~\text{where,}~T_i^\text{eff}=\frac{a_i^2}{\gamma r_i}.\label{eq_effect_thermal}
\ebq
Therefore the resetting wall at the boundary acts as an effective Langevin bath with effective temperature $T_i^\text{eff}$. In this high resetting rate, the average energy flux in the NESS Eq.~\eqref{eq:curreent} becomes identical to the thermally driven harmonic chain \cite{rllmothRieder} with higher order correction,
\bbq
J_\text{r}=\frac{k(T_1^\text{eff}-T_N^\text{eff})}{2 \gamma}\Bigg[ 1+\frac{m k}{2 \gamma^2}-\frac{m k}{2 \gamma^2}\sqrt{1+\frac{4 \gamma^2}{mk} } \Bigg]+O\Big(\frac{1}{r_j^2}\Big).
\ebq

Similarly, we can evaluate the two-time velocity correlation of a single oscillator and equal-time spatial velocity correlation in the limit of a high resetting rate. For large $r_i$, Eq.~\eqref{eq_reset_Cd} can be approximated as,
\bbq
\la v_l(t) v_{l}(0) \ra \simeq \frac{T_1^\text{eff}+T_N^\text{eff}}{2 m} J_0(\omega_c t).\label{thermal_Cd}
\ebq
Here $J_0(z)$ is the Bessel function of the first kind. Equal-time spatial velocity correlation in the limit of high resetting rate can be evaluated using Eqs.\eqref{limit_Cs2} and \eqref{limit_Cs} and the result is,
\bbq
\la v_l(t) v_{l'}(t) \ra \simeq \frac{T_1^\text{eff}+T_N^\text{eff}}{2 m} \delta_{ll'}.\label{thermal_Cs}
\ebq

In Sec.~\ref{sec_cross} we discuss the two-time velocity correlation of single oscillator and equal-time spatial velocity correlation for resetting driven oscillator chain. For the resetting driven chain, the nonzero value of $\la v_l(t) v_{l'}(t) \ra$ for $l \neq l'$ in the NESS implies that there exists a finite correlation between the velocity of the oscillators in the bulk. However, in the effective thermal limit, this correlation vanishes. Similarly, in the effective thermal limit, the two-time velocity correlation of single oscillator $\la v_l(t)v_l(0) \ra $ is given in Eq.~\eqref{thermal_Cd}. The small time behavior ($ t \ll \omega^{-1}_c $) of two-time velocity correlation $\la v_l(t)v_l(0) \ra$ is different for the resetting driven chain [Fig.~\ref{fig:correl}(a)] but the large time behavior is similar for both the cases.

\section{Conclusion}\label{sec_conclusion}

In this work, we study the NESS of a harmonic chain driven by resetting dynamics at the boundary. The harmonic chain is attached to two walls at the boundaries which undergo one-dimensional Brownian motion and stochastically reset to their initial position at different rates. As a result, the boundary oscillators experience exponentially correlated forces that lead to a NESS. For the resetting driven harmonic chain, we conclude from the numerical simulation that the velocity distributions of the bulk oscillators reach Gaussian distribution when the system size is thermodynamically large. The velocity distributions of the boundary oscillators are non-Gaussian and size-independent. We have numerically computed the velocity kurtosis profile and found that velocity kurtosis of the bulk oscillators decay as $N^{-1}$, where $N$ is the system size. We also show that the two-time velocity correlations of the bulk oscillators $\la v_l(t) v_{l}(0) \ra$ decay as $t^{-\frac{1}{2}}$ when $t \gg \omega_c^{-1}$. In the bulk, there exists a nonzero equal-time spatial velocity correlation $\la v_l(t) v_{l'}(t) \ra$ when $l \neq l'$. The distribution of the instantaneous energy current at the bulk is found to be size-dependent and reaches a stationary distribution only when $N \to \infty$. However, The distribution of instantaneous current at the boundary is size-independent and depends only on the resetting rate of the boundary walls.  It is important to note that, at a high resetting rate, an effective thermal picture emerges where we recover expression for the average energy current of a thermally driven harmonic chain. In this effective thermal limit, equal-time spatial velocity correlation $ \la v_l(t) v_{l'}(t) \ra$ vanishes when $l \neq l'$.  

In this paper, the boundary walls perform free diffusion with Poisson resetting where the waiting time between the two resetting events is drawn from an exponential distribution. However, the dynamics of the boundary walls can be generalized where the waiting time between two consecutive resetting events follow a power-law distribution \cite{particle_transport2,nonmarkovian_reset,resetting_review_int_particle} or the resetting mechanism becomes non-instantaneous \cite{noninstant_reset1,noninstant_reset2,noninstant_reset3,noninstant_reset4}. It is expected that such changes in the dynamics of the boundary walls will affect the NESS of the ordered harmonic chain and it is intriguing to explore the transport properties of such NESS. Another interesting question is how our results change when instead of our simple model of nonequilibrium bath, the boundary oscillators of the harmonic chain are attached to explicit nonequilibrium baths that are modeled using the stochastic resetting e.g. the model of nonequilibrium bath described in Ref.~\cite{noneq_bath}. It is also interesting to study how the energy transport in resetting driven oscillator chains gets affected if nonlinearity, mass disorder, and momentum non-conserving dynamics are introduced.

\section*{Acknowledgments}
The authors would like to thank Urna Basu and Ion Santra for useful discussions. R.S. acknowledges support from the Council of Scientific and Industrial Research, India [Grant No.
09/0575(11358)/2021-EMR-I].

\appendix

\section{Matrix formulation}\label{app_matrix}
In this appendix, we will briefly discuss the correlation calculations. To begin with, let us start with the matrix form the Langevin equations Eq.~\eqref{eq:eom},
\bbq
M\ddot{X}=-\Phi X(t) -\Gamma \dot{X}(t) + F(t),\label{eq_matrix_langevein}
\ebq
where $X(t)$ is a state vector of dimension $N\times 1$, and the $l$-th component is $x_l(t)$, displacement of the $l$-th oscillator from the equilibrium position.  $\Phi$ and $\Gamma$ are $N \times N$ matrix and the definitions of $\Phi$ and $\Gamma$ are,
\bbq
\Phi_{ij}=2 k \delta_{ij}-k \delta_{i,j-1}-k \delta_{i,j+1}\quad \text{and}\quad 
\Gamma_{ij}=\gamma\delta_{i1}\delta_{j1} +\gamma\delta_{iN}\delta_{jN}.
\ebq
$F(t)$ is the noise coming from the resetting wall. We can write these as,
\bbq
 F_j(t)=f_1(t)\delta_{j1}+f_N(t)\delta_{jN}.
\ebq
The correlation of this noise in the frequency domain is,
\bbq
\langle \tilde{F}(\omega)\tilde{F}^T(\omega')\rangle_{ij}=2\pi\delta(\omega+\omega')\left[ \tilde{g}(\omega,r_1)\delta_{i1}\delta_{j1}+\tilde{g}(\omega,r_N)\delta_{iN}\delta_{jN}\right], \label{eq_correl_reset}
\ebq
where $\tilde{g}(\omega,r_j)=\frac{2 a_j^2 r_j }{r_j^2+\omega^2}$. With the help of the Fourier transform, we can get rid of the time derivatives. The Fourier transform of the displacement matrix,
\bbq
\tilde{X}(\omega)=\int^{\infty}_{-\infty}dt e^{i\omega t}X(t).\label{eq_position_ft}
\ebq
Therefore the Langevin equation Eq.\eqref{eq_matrix_langevein} takes the form,
\bbq
\tilde{X}(\omega)=G(\omega) \tilde{F}(\omega)\quad\text{where}\quad G(\omega)=[-M\omega^2+\Phi-i\omega\Gamma]^{-1}. \label{eq_matrix_lang}
\ebq
$G(\omega)$ is the inverse of the tridiagonal matrix. The explicit form of this matrix is,
\bbq
 G(\omega)=\begin{bmatrix}
-m\omega^2+2k-i\omega \gamma & -k  &\cdots\\
-k & -m\omega^2+2k  & \cdots\\
\vdots & \ddots  & \cdots\\
0& \cdots &-m\omega^2+2k-i\omega \gamma \\ 
\end{bmatrix}^{-1}.
\ebq
$G$ is the inverse of a symmetric matrix, therefore $G$ it self is a symmetric matrix and $ G^*(\omega)=G(-\omega)$.
We can write the components of $G(\omega)$ \cite{tridiagonal}, 
\bbq
G_{l1}=(-k)^{l-1}\frac{\theta_{N-l}}{\theta_N}\quad\text{and}\quad
G_{lN}=(-k)^{N-l}\frac{\theta_{l-1}}{\theta_N}.\label{eq_G_matrix}
\ebq
where $\theta_l$ satisfy recursion relations,
\bbq
\theta_l = (-m \omega^2 +2k)\theta_{l-1}-k^2\theta_{l-2} \quad \forall l = 2,3,\cdots N-1,\label{eq_recursion}
\ebq 
and,
\bbq
 \theta_{N}=(-m\omega^2+2k-i\omega \gamma)\theta_{N-1}-k^2\theta_{N-2},
\ebq
\newline
with the boundary condition,
\bbq
\theta_0 = 1,\qquad \theta_1 = -m\omega^2 + 2k-i\omega\gamma .
\ebq
Solving the recursion relations of Eq.~\eqref{eq_recursion} we get,
\bbq
\theta_l = \frac{(-k)^{l-1}}{\sin{q}}[k \sin{(l+1)q}-i\omega \gamma \sin{lq}]\quad\text{and}\quad \theta_N = \frac{(-k)^N}{\sin{q}}[a(q)\sin{Nq}+b(q)\cos{Nq}]. 
\ebq
Here,
\bbq
a(q) = -\frac{2i\gamma \omega}{k}+\cos{q}\left(1-\frac{\gamma^2 \omega^2}{k^2} \right) \quad\text{and}\quad b(q) = \sin{q}\left( 1-\frac{\gamma^2 \omega^2}{k^2}\right),
\ebq
and $\omega$ and $q$ are related through,
\bbq
\cos{q}=\left(1-\frac{m\omega^2}{2k}\right),\quad \text{and}\quad \omega= \omega_c \sin{\frac q2}. \label{eq:omega}
\ebq
with $\omega_c=2\sqrt{k/m}$.

\section{Calculation for spatio-temporal two-point correlation of velocity}\label{app_reset}
In this section, we will calculate the spatio-temporal two-point correlation of the velocity of $l$-th oscillator $ \la v_l(t)v_{l'}(t') \ra$ in the steady state for resetting driven oscillator chain. The velocity of the $l$-th oscillator of the resetting driven chain can be written using Eqs.~\eqref{eq_position_ft} and \eqref{eq_matrix_lang},
\bbq
v_l(t)=\int_{-\infty}^{\infty} \frac{d \omega}{2 \pi}e^{-i\omega t}(-i \omega)\big[G_{l1}(\omega)\tilde{f}_1(\omega)+G_{lN}(\omega)\tilde{f}_N(\omega)\big].
\ebq
Here $\tilde{f}_j(\omega)$ is the Fourier transform of the resetting force, $f_j(t)$. Using Eq.~\eqref{eq_correl_reset}, we can write the spatio-temporal correlation of the velocity as,
\bbq
\la v_l(t)v_{l'}(t') \ra = \int_{-\infty}^{\infty} \frac{d\omega}{2 \pi} \omega^2 e^{-i\omega (t-t')}\big[G_{l1}(\omega)G_{l'1}^*(\omega)\tilde{g}(\omega,r_1)+G_{lN}(\omega)G_{l'N}^*(\omega)\tilde{g}(\omega,r_N)\big].\label{eq_cross_v_reset}
\ebq
\subsection{Two-time velocity correlation of single oscillator:}\label{appendix_two_time_correl_reset} For $l=l'$ and $t>t'$,
\begin{equation}
\la v_l(t) v_{l}(t') \ra=\int_{-\infty}^{\infty} \frac{d\omega}{2 \pi} \omega^2 e^{-i\omega (t-t')}\big[|G_{l1}(\omega)|^2\tilde{g}(\omega,r_1) +|G_{lN}(\omega)|^2 \tilde{g}(\omega,r_N)\big].
\end{equation}
From the above equation, it is clear that the two-time velocity correlation of single oscillator $\la v_l(t) v_{l}(t') \ra$ for resetting driven chain, can be written as a sum of two separate contributions coming from two resetting walls at the two boundaries, i.e.,
\bbq
\la v_l(t) v_{l}(t') \ra=\bar{C}(r_1,t,t')+\bar{C}(r_N,t,t').
\ebq
The contribution from the left resetting wall,
\bbq
\bar{C}(r_1,t,t')=\frac{1}{2\pi}\int_{-\infty}^{\infty} d\omega \omega^2 e^{-i\omega (t-t')}|G_{l1}(\omega)|^2\tilde{g}(\omega,r_1).
\ebq
$\bar{C}(r_1,t,t') $ will have a non-zero contribution from the even terms with respect to $\omega$. Using Eq.~\eqref{eq_G_matrix} and keeping only the terms that are even in $\omega$, we get,
\begin{eqnarray}
\bar{C}(r_1,t,t') &=& \frac{1}{k^4}\int_{0}^{\infty} \frac{d\omega}{\pi} \omega^2 \cos{[\omega (t-t')]}
\Bigg(\frac{k^2 \sin^2{(N q-l q+q)}}{|a(q)\sin{Nq}+b(q)\cos{Nq}|^2}\cr
&+&\frac{\omega^2 \gamma^2 \sin^2{(N q-l q)}}{|a(q)\sin{Nq}+b(q)\cos{Nq}|^2}  \Bigg)\tilde{g}(\omega,r_1)
\end{eqnarray}

We are interested in the time correlation of velocity in the bulk of the chain, therefore we take $l=N/2+\epsilon$, and take the limit $\epsilon \ll N$, which leads to,
\begin{eqnarray}
\bar{C}(r_1,t,t')&=& \frac{1}{k^4}\int_{0}^{\infty} \frac{d\omega}{2 \pi} \omega^2 \cos{[\omega (t-t')]}\Bigg(\frac{(k^2+\omega^2 \gamma^2)}{|a(q)\sin{Nq}+b(q)\cos{Nq}|^2}\cr
&-&\frac{ k^2 \cos{(Nq -2\epsilon q +2q)}+\omega^2 \gamma^2 \cos{(Nq-2 \epsilon q)} }{|a(q)\sin{Nq}+b(q)\cos{Nq}|^2}\Bigg) \tilde{g}(\omega,r_1).\label{ap_thermod_limit}
\end{eqnarray}
For $\omega>\omega_c$, $q$ becomes complex. In the thermodynamically large system size, the integrand vanishes exponentially as $e^{-2N\bar{q}}$ in the region $\omega>\omega_c$, here $\bar{q}$ is real. The range of the integration reduces to $0\leq\omega\leq\omega_c$, or $0\leq q\leq \pi$. In the large-$N$ limit, $\cos{N q}$ is a highly oscillatory function and we can evaluate these integrations by averaging over fast oscillation in $x=N q$, using the following identities (see 2.558 in \cite{integral_table}),
\begin{eqnarray}
\frac{1}{2 \pi} \int^{2\pi}_0\frac{d x}{(c_1 \sin{x}+d \cos{x})^2+c_2^2 \sin^2{x}}&=&-\frac{1}{c_2 d},~\text{for}~c_2<0,\n\\
\frac{1}{2 \pi} \int^{2\pi}_0\frac{d x\cos{x}}{(c_1 \sin{x}+d \cos{x})^2+c_2^2 \sin^2{x}}&=&0. \label{ap:fastos}
\end{eqnarray}
with $c_2=\text{Im}[a(q)]=-\frac{2\gamma \omega}{k}$ and $d=\text{Re}[b(q)]=\sin{q}\left( 1+\frac{\gamma^2 \omega^2}{k^2}\right)$. Therefore Eq.~\eqref{ap_thermod_limit} can be written as,
\begin{eqnarray}
\bar{C}(r_1,t,t')= \frac{1}{k^4}\int_{0}^{\infty} \frac{d\omega}{2\pi} \omega^2 \cos{[\omega (t-t')]}\frac{k^2+\gamma^2\omega^2}{-d c_2}\tilde{g}(\omega,r_1).
\end{eqnarray}
Finally, we arrive at,
\begin{eqnarray}
\bar{C}(r_1,t,t') &=& \frac{1}{k} \int_{0}^{\pi} \frac{dq}{\pi} \Big|\frac{d \omega}{d q} \Big|\frac{\omega \cos{[\omega (t-t')]}}{4\gamma \sin{q}}\tilde{g}(\omega,r_1)\cr
&=&\frac{a_1^2 r_1}{\gamma m}\int_{0}^{\pi} \frac{dq}{2\pi} \frac{\cos \Big[\omega_c \sin{\Big(\frac{q}{2}\Big)}(t-t') \Big] }{r_1^2+\omega_c^2 \sin^2{\frac{q}{2}}}
=a_1^2 C(r_1,t,t'),\label{eq_correl_time_part}
\end{eqnarray}
where we have used the explicit form of $\tilde{g}(\omega,r_1)$ and $\omega(q)$ from Eq.~\eqref{eq_correl_reset} and \eqref{eq:omega}. Similarly, we can evaluate the contribution from the right resetting wall, and by combining these two results we arrive at the,
\bbq
\la v_l(t) v_{l}(t') \ra=\frac{1}{ \gamma m}\int_{0}^{\pi} \frac{dq}{2 \pi} \cos \Big[\omega_c \sin{\Big(\frac{q}{2}\Big)}(t-t') \Big]\Bigg[\frac{a_1^2 r_1 }{r_1^2+\omega_c^2 \sin^2{\frac{q}{2}}}+\frac{a_N^2 r_N }{r_N^2+\omega_c^2 \sin^2{\frac{q}{2}}}\Bigg].\label{eq_correl_time}
\ebq
This integral can be evaluated numerically.

\noindent \textbf{Large time behavior ($t-t'\gg \omega_c^{-1}$):}
To calculate the large time behavior of $\la v_l(t) v_{l}(t') \ra$, we first substitute $z=\omega_c \sin \Big(
\frac{q}{2}\Big)$ in Eq.~\eqref{eq_correl_time_part} and arrive at,
\bbq
 C(r_1,t,t')=\frac{ r_1}{\gamma m}\int_{0}^{\omega_c}\frac{d z}{\pi} \frac{\cos \big(z (t-t')\big)}{\sqrt{\omega_c^2-z^2}(r_1^2+z^2)}.
\ebq 
For large $t-t'$, $\cos \big(z (t-t')\big)$ is a bounded and fast oscillatory function. Therefore, the dominating contribution to the integral is coming from the region $z=\omega_c$ and contributions from small $z$ are negligible. We can now approximate $r_1^2+z^2$ as $r_1^2+\omega_c^2$. Therefore the last integral becomes,
\bbq
 C(r_1,t,t')\simeq \frac{ r_1}{\gamma m}\int_{0}^{\omega_c}\frac{d z}{\pi} \frac{\cos \big(z (t-t')\big)}{\sqrt{\omega_c^2-z^2}(r_1^2+\omega_c^2)}=\frac{ r_1}{2 \gamma} \frac{J_0\big(\omega_c (t-t')\big)}{(m r_1^2+4 k)}\label{eq_time_part_large}.
\ebq
Here $J_0(z)$ is the Bessel function of the first kind. Therefore when $t-t'$ is large, using Eqs.~\eqref{eq_correl_time_part} and \eqref{eq_time_part_large} we get
\bbq
\la v_l(t) v_l(t') \ra \simeq\Bigg[\frac{a_1^2 r_1}{2 \gamma (mr_1^2+4 k)}+\frac{a_N^2 r_N}{2 \gamma (mr_N^2+4 k)} \Bigg]J_0\big(\omega_c (t-t')\big).
\ebq

\noindent \textbf{Kinetic temperature profile at the bulk:} The bulk value of the kinetic temperature profile can be evaluated using Eq.~\eqref{eq_correl_time} by taking $t=t'$. Therefore, in the stationary state,
\bbq
\la v_l(t)^2 \ra =\frac{1}{ \gamma}\int_{0}^{\pi} \frac{dq}{2 \pi} \Bigg[\frac{a_1^2 r_1 }{mr_1^2+4k \sin^2{\frac{q}{2}}}+\frac{a_N^2 r_N }{mr_N^2+4k \sin^2{\frac{q}{2}}}\Bigg].
\ebq
The last integral can be evaluated exactly, which leads to the following expression of the bulk value of the kinetic temperature profile,
\bbq
\hat{T}_\text{bulk}=m\la v_l^2(t) \ra=\frac{a_1^2 }{2 \gamma\sqrt{r_1^2+\frac{4 k }{m}}}+\frac{a_N^2 }{2 \gamma\sqrt{r_N^2+\frac{4 k }{m}}}.
\ebq

\subsection{Equal-time spatial velocity correlation:}\label{appendix_eq_time_reset}

For $l\neq l'$ and $t=t'$, using Eq.~\eqref{eq_cross_v_reset},
\begin{equation}
\la v_l(t)v_{l'}(t) \ra = \int_{-\infty}^{\infty} \frac{d\omega}{2 \pi} \omega^2 \big[G_{l1}(\omega)G_{l'1}^*(\omega)\tilde{g}(\omega,r_1)+G_{lN}(\omega)G_{l'N}^*(\omega)\tilde{g}(\omega,r_N)\big].
\end{equation}
Similar to the previous calculation, $\la v_l(t)v_{l'}(t) \ra$ can be written as a sum of two separate contributions coming from resetting walls at the boundaries as the following,
\bbq
\la v_l(t)v_{l'}(t) \ra =\bar{\theta}(r_1,l,l')+\bar{\theta}(r_N,l,l').
\ebq
The contribution from the left resetting wall,
\bbq
\bar{\theta}(r_1,l,l')=\int_{-\infty}^{\infty} \frac{d\omega}{2\pi} \omega^2 G_{l1}(\omega)G_{l'1}^*(\omega)\tilde{g}(\omega,r_1).
\ebq
Using Eq.~\eqref{eq_G_matrix} and keeping only the terms that are even in $\omega$, we get,
\begin{eqnarray}
\bar{\theta}(r_1,l,l')&=&\frac{1}{ k^4}\int_{0}^{\infty} \frac{d\omega}{2\pi} \omega^2 \Bigg( \frac{(k^2+\gamma^2 \omega^2)\cos{(l' q-lq)}}{|a(q)\sin{Nq}+b(q)\cos{Nq}|^2}\cr
&-&\frac{k^2 \cos{(Nq-\epsilon q +2 q)+\omega^2 \gamma^2 \cos{(N q-\epsilon q)}}}{|a(q)\sin{Nq}+b(q)\cos{Nq}|^2}\Bigg)\tilde{g}(\omega,r_1).
\end{eqnarray}
Now we are interested in $l=\frac{N}{2}$, $l'= \frac{N}{2}+\alpha$ where $\alpha$ is an integer. For $\omega>\omega_c$, the integrand vanishes [see the discussion after Eq.~\eqref{ap_thermod_limit}]. The final expression after averaging over fast oscillation using Eq.~\eqref{ap:fastos} is,
\bbq
\bar{\theta}(r_1,l,l')=\frac{1}{ k^4}\int_{0}^{\pi} \frac{dq}{2\pi} \Big| \frac{d\omega}{dq} \Big| \omega^2 \frac{(k^2+\gamma^2 \omega^2)\cos{(l'-l)q}}{-d c_2}\tilde{g}(\omega,r_1).
\ebq
Using  $c_2=\text{Im}[a(q)]=-\frac{2\gamma \omega}{k}$ and $d=\text{Re}[b(q)]=\sin{q}\left( 1+\frac{\gamma^2 \omega^2}{k^2}\right)$, and explicit form of $\tilde{g}(\omega,r_1)$, we arrive at,
\bbq
\bar{\theta}(r_1,l,l')=\frac{a_1^2 r_1^2}{ \gamma}\int_{0}^{\pi} \frac{dq}{2\pi} \frac{\cos{(l'q-lq)}}{m r_1^2+4 k \sin^2\frac{q}{2}}
= \frac{a_1^2 }{2\gamma} \theta(r_1,l,l') ,
\ebq
where, 
\begin{equation}
\theta(r_i,l,l')=\frac{r_i}{m r_i^2+4 k}\,{}_3 \tilde{F}_2\,\Bigg[1/2,1,1;1-l'+l,1+l'-l;\frac{4 k}{4 k+mr_i^2} \Bigg].
\end{equation}
Here $l'-l$ is an integer and ${}_p \tilde{F}_q$ is a generalized regularized hypergeometric function~\cite{DLMF}. The definition of generalized regularized hypergeometric function ${}_p \tilde{F}_q$ is,
\begin{equation}
{}_p \tilde{F}_q[x_1,x_2\cdots x_p;y_1,y_2,\cdots y_q;z]=\frac{{}_p {F}_q[x_1,x_2\cdots x_p;y_1,y_2,\cdots y_q;z]}{\Gamma(y_1)\Gamma(y_2)\cdots\Gamma(y_q)}, 
\end{equation}
where $\Gamma(z)$ is the gamma function and ${}_p {F}_q$ is the generalized hypergeometric function. The definition of generalized hypergeometric function ${}_p {F}_q$ is,
\begin{equation}
{}_p {F}_q[x_1,x_2\cdots x_p;y_1,y_2,\cdots y_q;z]=\sum_{j=0}^{\infty}\frac{(x_1)_j(x_2)_j\cdots (x_p)_j}{(y_1)_j(y_2)_j\cdots (y_q)_j}\Big(\frac{z^j}{j!}\Big).
\end{equation}
$(x)_j$ is the Pochhammer symbol defined as $(x)_j\equiv \Gamma(x+j)/\Gamma(x)$ \cite{DLMF}.
We can calculate the contribution coming from the right resetting wall $\bar{\theta}(r_N,l,l') $ in a similar manner. Combining both contributions we arrive at the,
\begin{equation}
\la v_l(t) v_{l'}(t) \ra = \frac{1}{2 \gamma} \Big[ a_1^2 \theta(r_1,l,l')+a_N^2 \theta(r_N,l,l') \Big].
\end{equation}

\end{document}